\documentclass[pra,twocolumn,floatfix,superscriptaddress,amsmath,citeautoscript,aps,longbibliography]{revtex4-1}
\usepackage{graphicx}  
\usepackage{dcolumn}   
\usepackage{bm}        
\usepackage{amssymb}   
\usepackage{amsmath}
\usepackage{color}
\usepackage{natbib}
\usepackage{hyperref}  
\hypersetup{hypertex=true,
            colorlinks=true,
            linkcolor=blue,
            anchorcolor=blue,
            citecolor=blue}

\newcommand{\KBGBO}{KBaGd(BO$_3$)$_2$}

\begin{document}


\title{Magnetic phase diagrams and large magnetocaloric effects of the two-dimensional antiferromagnetic triangular lattice of Gd$^{3+}$ ions in KBaGd(BO$_3$)$_2$}

\date{\today}
\author{Z.~M. Song}
\thanks{These authors have contributed equally to this work.}
\affiliation{Department of Physics, Southern University of Science and Technology, Shenzhen 518055, China}

\author{N. Zhao}
\thanks{These authors have contributed equally to this work.}
\affiliation{Department of Physics, Southern University of Science and Technology, Shenzhen 518055, China}

\author{H. Ge}
\affiliation{Department of Physics, Southern University of Science and Technology, Shenzhen 518055, China}

\author{T.~T. Li}
\affiliation{Department of Physics, Southern University of Science and Technology, Shenzhen 518055, China}

\author{J. Yang}
\affiliation{Department of Chemsitry, Southern University of Science and Technology, Shenzhen 518055, China}

\author{L. Wang}
\affiliation{Department of Physics, Southern University of Science and Technology, Shenzhen 518055, China}
\affiliation{Shenzhen Institute for Quantum Science and Engineering, Southern University of Science and Technology, Shenzhen 518055, China}

\author{Y. Fu}
\affiliation{Department of Physics, Southern University of Science and Technology, Shenzhen 518055, China}

\author{Y. Z. Zhang}
\affiliation{Department of Chemsitry, Southern University of Science and Technology, Shenzhen 518055, China}

\author{S. M. Wang}
\affiliation{Department of Physics, Southern University of Science and Technology, Shenzhen 518055, China}

\author{J.~W. Mei}
\affiliation{Department of Physics, Southern University of Science and Technology, Shenzhen
  518055, China}
\affiliation{Shenzhen Institute for Quantum Science and Engineering, Southern University of Science and Technology, Shenzhen 518055, China}
\affiliation{Shenzhen Key Laboratory of Advanced Quantum Functional Materials
   and Devices, Southern University of Science and Technology, Shenzhen 518055, China}

\author{H. He}
\affiliation{School of Chemical Science, University of Chinese Academy of Sciences (UCAS), Beijing 100190, China}

\author{S. Guo}
\affiliation{Shenzhen Institute for Quantum Science and Engineering, Southern University of Science and Technology, Shenzhen 518055, China}
\affiliation{International Quantum Academy, Shenzhen 518048, China.}

\author{L.~S. Wu}
\email{wuls@sustech.edu.cn}
\affiliation{Department of Physics, Southern University of Science and Technology, Shenzhen 518055, China}
\affiliation{Shenzhen Key Laboratory of Advanced Quantum Functional Materials and Devices, Southern University of Science and Technology, Shenzhen 518055, China}
\affiliation{Quantum Science Center of Guangdong-Hong Kong-Macao Greater Bay Area (Guangdong), Shenzhen 518045, China}

\author{J.~M. Sheng}
\email{shengjm@sustech.edu.cn}
\affiliation{Department of Physics, Southern University of Science and Technology,
Shenzhen 518055, China}
\affiliation{Academy for Advanced Interdisciplinary Studies, Southern University of Science and Technology,
Shenzhen 518055, China}

\date{\today}

\begin{abstract}
We report a detailed study of the magnetic properties of KBaGd(BO$_3$)$_2$, in which magnetic Gd$^{3+}$ ($S=7/2$) ions form into two-dimensional triangular layers. Magnetization, specific heat and magnetocaloric effect (MCE) measurements have been performed on KBaGd(BO$_3$)$_2$ single crystals. The results show that a long-range antiferromagnetic state is established below $T_{\rm N}=0.24$ K. In zero fields, only about half of the full entropy is released at $T_{\rm N}$, indicating that not all the magnetic moments are frozen below the ordering temperature, as expected from the geometrical frustration of the triangular spin lattice. Further studies under external fields were performed down to 50 mK, and the magnetic phase diagrams are established with magnetic fields applied both within and perpendicular to the triangular plane. KBaGd(BO$_3$)$_2$ serves as an example of a two-dimensional triangular lattice with large spin values ($S=7/2$) and can be directly compared with the iso-structure KBaR(BO$_3$)$_2$ (R = Dy-Yb) family of doublet ground states, which exhibit effective spins of $S=1/2$. 
\end{abstract}

\maketitle

\section{INTRODUCTION}
Frustrated spin systems are one of the most attractive manifestations in demonstrating exotic collective phenomena in condensed matter physics~\cite{Ramirez1994,Moessner2006,Balents2010}. Geometrical spin frustration arises when the system cannot realize a spin configuration where all the interactions are minimized simultaneously, and thus cannot be stable at a unique ground state. As a consequence, strong quantum fluctuations are induced, and macroscopic ground state degeneracies with various exotic physics emerge at low temperatures~\cite{Ramirez1994,Harris1997,Han2012,Zapf2014,Starykh2015}. As a prototype example of the realization of geometrical frustration, the triangular spin lattices with antiferromagnetic (AFM) exchange interactions have been well explored in the past~\cite{Miyashita1986,Chubukov1991,SY2008,YS2016,SI2017,MM2019,Sheng2022}. Considering only the nearest-neighbor Heisenberg AFM exchange interactions, the magnetic ground state of the triangular spin lattice is proposed to be an ordered state of $120^{\rm o}$ spin configuration with a significantly reduced staggered moments due to the large quantum spin fluctuations~\cite{Huse1988,Jolicoeur1989,Singh1992,Bernu1994,White2007}.

Up to now, most studies on triangular lattices have been focused on systems with spins $S=1/2$~\cite{Coldea2001,Shirata2012,YS2015,MM2019,Sheng2022}, where the quantum fluctuations are expected to be the greatest. In general, the strength of the quantum fluctuations will be reduced with larger values of $S$. However, experimental studies of the triangular spin lattices with different spin values are still relatively rare. The recently discovered rare earth borate family compounds KBaR(BO$_3$)$_2$ with R=Y, Gd, Tb, Dy, Ho, Tm, Yb and Lu, have provided us a unique opportunity~\cite{SG2019}. In this family of materials, the magnetic rare earth ions form into perfect two-dimensional edge shared triangular lattices, with the non-magnetic potassium (K) and barium (Ba) ions stuffed in between. For rare earth ions such as Yb, an effective $S=1/2$ ground state is usually realized due to the crystalline electric field (CEF) splitting. On the other hand, the orbital contribution of Gd$^{3+}$ is very small and a nearly isotropic $S=7/2$ ground state is expected. Previous studies of KBaGd(BO$_3$)$_2$ reveal antiferromagnetic interactions between these magnetic Gd$^{3+}$ ions based on the magnetization measurements~\cite{SG2019}. In addition, no long-range magnetic order was found above 1.8 K~\cite{SG2019}. It is still unclear whether the ground state is ordered below 1.8 K, or whether a liquid or glass-like order is built up instead, as in the frustrated gadolinium garnet system Gd$_3$Ga$_5$O$_{12}$~\cite{OA1998,OA1999,YK2001,OA1997}.

In this paper, we have performed the low temperature magnetization and specific heat measurements on KBaGd(BO$_3$)$_2$ single crystals. Long-range magnetic order is observed at $T_{\rm N}=0.24$ K in zero field. In addition, this order is gradually suppressed with the application of a magnetic field both along and perpendicular to the triangular lattice. No fractional plateau-like phases are induced in magnetic fields in either direction. On the other hand, only half of the full entropy is released at the transition temperature, indicating that a large amount of fluctuations persist above the transition temperature.



\section{EXPERIMENTAL DETAILS}
Single crystals of KBaGd(BO$_3$)$_2$ were synthesized, following the flux method reported in~\cite{SG2019}. Pure raw materials of K$_2$CO$_3$, BaCO$_3$, and Gd$_2$O$_3$ were mixed with stoichiometric ratio, while additional H$_3$BO$_3$ and KF were used as flux. All these precursors were placed in a platinum crucible. The temperature profile as described in~\cite{SG2019} was used, and transparent plate-like single crystals of KBaGd(BO$_3$)$_2$ with sizes about $1\times1\times0.5$ $\rm mm^3$ were obtained after removing the left flux with distilled water.

The crystal structure and orientations of KBaGd(BO$_3$)$_2$ were verified using a Bruker D8 Quest diffractometer with Mo-K$\alpha$ radiation ($\lambda$=0.71073 $\AA$). All the thermal property measurements were performed on single crystals, using the commercial Quantum Design Physical Property Measurement System (PPMS) and Magnetic Property Measurement System (MPMS).

\section{Results and Analysis}

\subsection{Crystal Structure}
\begin{figure}[t!]
 \includegraphics[width=2.7in]{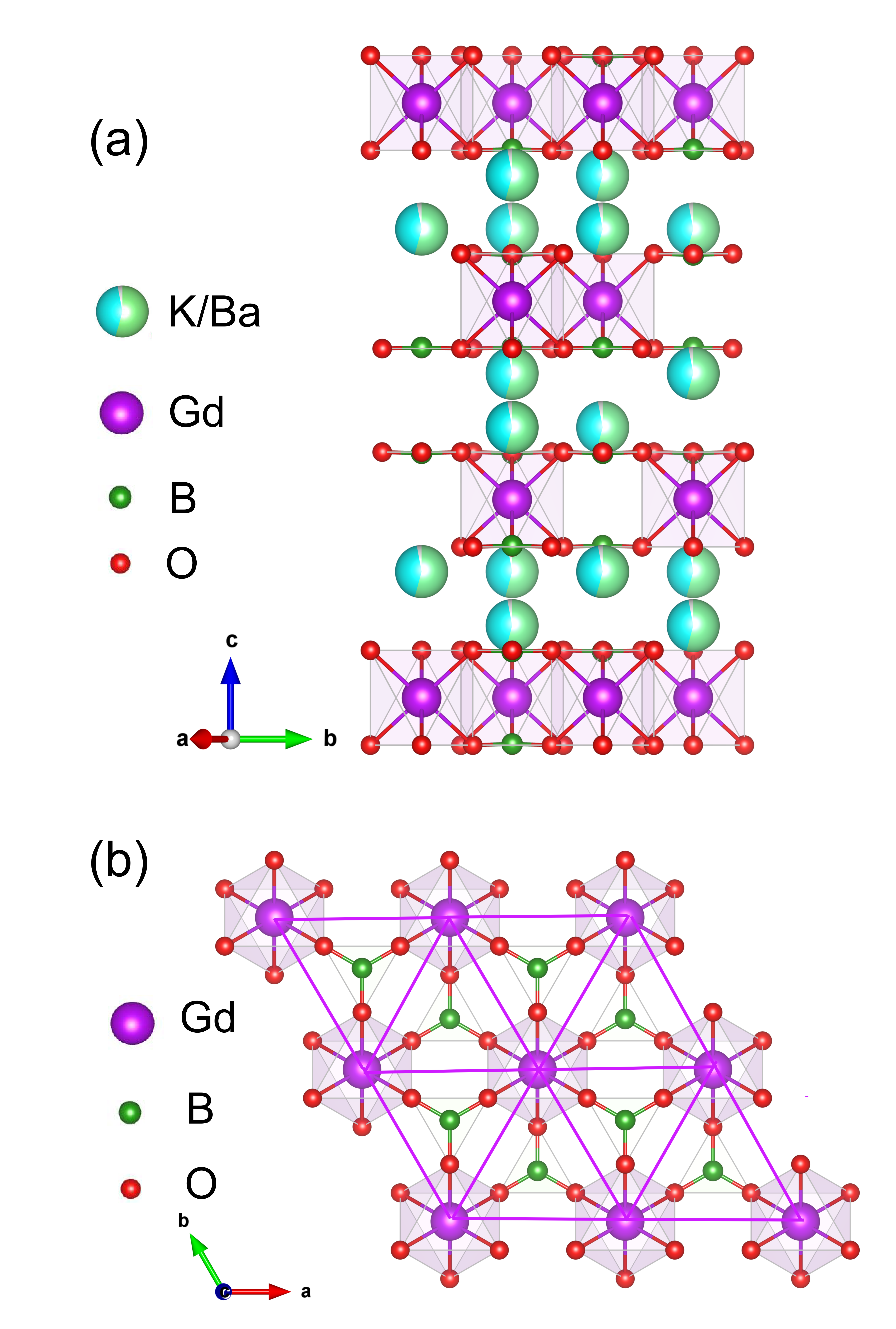}
    \caption{(a) The crystal structure of KBaGd(BO$_3$)$_2$, where Gd$^{3+}$ ions form into a perfect triangular layer in the $ab$ plane and these triangular layers are stacked along the $c$ axis. (b) Top view of the triangular layer of GdO$_6$ octahedra.}
\label{fig:structure}
\end{figure}

\KBGBO~crystallized in the hexagonal structure with the space group $R\bar{3}$\textit{m} (Fig.~\ref{fig:structure}(a)). The single crystal structure has been verified at 100 K by the single crystal X-ray diffractometer, and the details of the refinement are given in Table \ref{tab:structure1} and \ref{tab:structure2}. The refined lattice parameters are found out to be $a=b=5.4638(2)$ {\AA}, $c=17.9461(3)$ {\AA} with $\alpha=\beta=90^{\rm o}$ and $\gamma=120^{\rm o}$, which are consistent with previous reports~\cite{SG2019}. In~\KBGBO, the magnetic Gd$^{3+}$ ions form into an ideal triangular layer in the $ab$-plane, with the nearest-neighbor Gd-Gd distance about 5.4638 {\AA} (Fig.~\ref{fig:structure}(b)). These magnetic layers are then stacked along the $c$ axis, with an interlayer distance of about 5.982 {\AA}. Non-magnetic ions such as K$^+$ and Ba$^{2+}$ are stuffed in between these two-dimensional Gd layers. Our single crystal X-ray refinement shows that these K$^+$ and Ba$^{2+}$ ions share the same crystallographic site ($6c$), and are randomly distributed with a ratio close to $0.510(5):0.495(5)\approx1:1$ (Table~\ref{tab:structure2}). Similar site disorder effects are also found in the isostructural compounds KBa$R$(BO$_3$)$_2$, where $R$=Y, Gd, Tb, Dy, Ho, Tm, Yb and Lu~\cite{SG2019,MB2017}.

The site disorder impact on the magnetic properties of these rare earth ions are still under debate. Similar situations have been found in the rare earth triangular lattice magnet YbMgGaO$_4$~\cite{YS2015,YSLD2017}, where site disorder was found between the non-magnetic Mg$^{\rm 2+}$ and Ga$^{\rm 3+}$ ions. Continuum-like spin excitations were observed in the inelastic neutron scattering experiments~\cite{Paddison2017,YS2016}. Based on this, spin liquid ground states with spinon fermi surfaces have been proposed~\cite{YS2016,YDL2017}. However, it was also argued that the effects of disorder play a more important role~\cite{ZM2018,Rao2021}. For the rare earth elements with $L\neq0$, the ground states are determined by the local charge environments due to the CEF splitting. The presence of the site disorder with different charges leads to distinct ground states on different sites. Thus, the intrinsic properties of YbMgGaO$_4$ might be affected by the disorder of the Mg$^{\rm 2+}$ and Ga$^{\rm 3+}$ ions. Unlike Yb$^{3+}$ ions, Gd$^{3+}$ ions of the ground state of $^8S_{7/2}$ ($S=7/2, L=0$) have much smaller orbital contributions. Weak single-ion anisotropy may arise from the mixing of the higher-level states at low temperatures~\cite{Glazkov2005,Glazkov2006}. Even with this considered, the CEF effect of Gd$^{3+}$ and the orbital contribution are still much weaker than other rare-earth ions. Thus, the influence of the disorder of K$^{\rm +}$ and Ba$^{\rm 2+}$ ions is minimized in KBaGd(BO$_3$)$_2$.

In addition, the distances between the magnetic rare-earth ions in the borate family KBaR(BO$_3$)$_2$ are in the range of about $5\sim6$ \AA. This usually results in very weak spin correlations. However, due to the large spin value $S=7/2$, the exchange interactions for Gd$^{3+}$ ions are greatly enhanced, as compared to other family compounds such as KBaYb(BO$_3$)$_2$~\cite{Tokiwa2021,Pan2021}.

\begin{table}[h]
\caption{Refined parameters and crystallographic data of~\KBGBO~from single crystal X-ray diffraction pattern.}
\centering
\begin{tabular}{p{5.2cm}<{\raggedright}p{3cm}<{\raggedright}}
    \hline
    ~~~~Chemical Formula                 & \KBGBO\\
    \hline
    ~~~~Formula Weight                   & 451.295\\
    ~~~~Temperature(K)                   & 100\\
    ~~~~$\lambda(\rm\AA)$                   & 0.71073\\
    ~~~~Crystal system                   & Trigonal\\
    ~~~~Space group, $\textit{Z}$        & $R\bar{3}$\textit{m}(No.166), 3\\
    ~~~~$a(\rm\AA)$                           & 5.4638(2)\\
    ~~~~$b(\rm\AA)$                           & 5.4638(2)\\
    ~~~~$c(\rm\AA)$                           & 17.9461(3)\\
    ~~~~$\alpha(\rm^{\circ})$    & 90\\
    ~~~~$\beta(\rm^{\circ})$    & 90\\
    ~~~~$\gamma(\rm^{\circ})$    & 120\\
    ~~~~$V(\rm\AA^{3})$                       & 464.0\\
    ~~~~$\rho_{\rm calc}(\rm g/cm^{3})$           & 4.85\\
    ~~~~Goodness-of-fit                    & 3.28\\
    ~~~~$R_{1}$                            & 5.54\\
    ~~~~$wR_{2}$                           & 6.12\\

\hline 

\end{tabular}

\label{tab:structure1}
\end{table}

\begin{table}[h]
\caption{Atomic displacement parameters of \KBGBO.}
\centering 
\begin{tabular}{p{0.8cm}<{\centering}p{0.6cm}<{\centering}p{1.2cm}<{\centering}p{1.2cm}<{\centering}p{1.2cm}<{\centering}p{1.4cm}<{\centering}p{1.1cm}<{\centering}}
\hline
Atom & Wkf. & x  & y & z & Occupancy & U$_{eq}$\\
\hline
Ba    & 6c  & 0 & 0 & 0.7877(3) & 0.510(5) &0.015(7)\\
K   & 6c  & 0 & 0 & 0.7876(9) & 0.495(5) &0.015(9)\\
Gd   & 3a  & 0 & 0 & 0 & 1 &0.009(4)\\
B   & 6c  & 0 & 0 & 0.590(6) & 1 &0.010(8)\\
O   & 18h  & 0.291(2) & 0.145(9) & 0.413(1)  & 1 &0.015(6)\\
\hline 
\end{tabular}

\label{tab:structure2}
\end{table}

\subsection{Magnetization and Phase Diagram}
\begin{figure}[ht!]
 \includegraphics[width=2.7in]{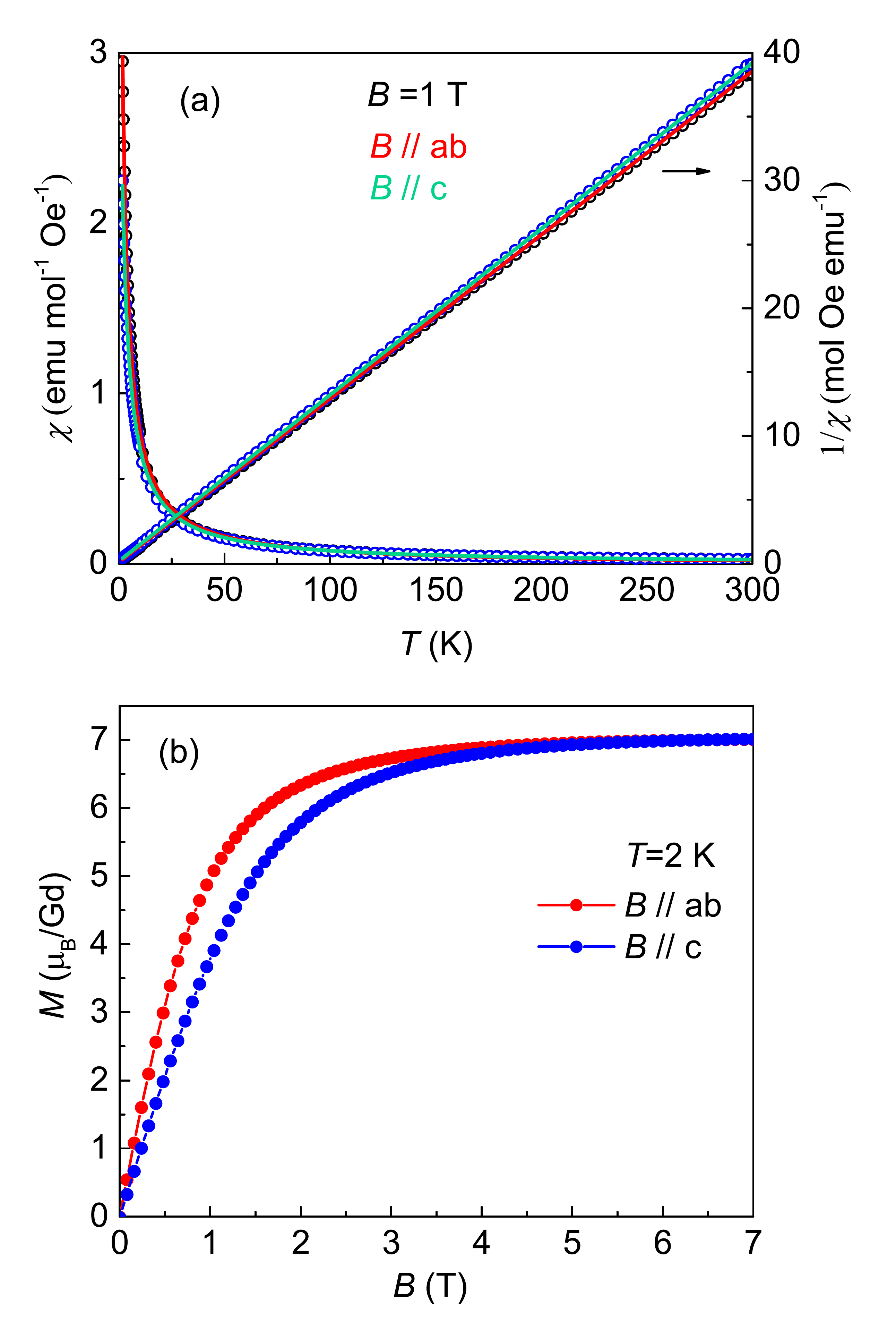}
    \caption{(a) Temperature dependence of the magnetic susceptibility $\chi$ (left axis) and the inverse magnetic susceptibility 1/$\chi$ (right axis) with the magnetic field parallel to the $ab$ plane and $c$ axis from 2-300 K. The blue and red solid lines indicate a Curie-Weiss fit to the inverse magnetic susceptibility over the entire temperature range.  (b) Isothermal magnetization $M$(B) measured at 2 K with the magnetic field applied in the $ab$ plane and $c$ axis. }
\label{fig:magnetization}
\end{figure}

The temperature-dependent dc magnetic susceptibility $\chi(T)$ with field applied in the $ab$-plane and $c$ axis was measured, as shown in Fig.\ref{fig:magnetization}(a). A weak anisotropy is observed for the temperature range $2-300$ K, and no sign of long-range magnetic order is observed for both directions above 2 K. The blue and red solid lines in Fig.\ref{fig:magnetization}(a) are the fits of the inverse magnetic susceptibility to the Curie-Weiss law, $\chi= C/(T- \theta\rm_{CW}$), where $\theta\rm_{CW}$ is the Curie-Weiss temperature and $C$ is a constant associated with the effective magnetic moment. The obtained Curie-Weiss temperatures are $\theta\rm_{CW}^{ab}= -1.1\pm0.2$ K and $\theta\rm_{CW}^{c}= -1.3\pm0.2$ K and the effective magnetic moments are $\mu{\rm_{eff}^{ab}}$=7.9 $\mu\rm_B$ and $\mu{\rm_{eff}^{c}}$=7.8 $\mu\rm_B$, for both field directions respectively. These negative Curie-Weiss temperatures indicate weak antiferromagnetic exchange interactions in KBaGd(BO$_3$)$_2$, similar to the observations in~\cite{SG2019}.

 Isothermal magnetization $M$(B) was measured at 2 K as presented in Fig.~\ref{fig:magnetization}(b). The measured saturation moments are about 7 $\mu\rm_B$/Gd, for both field directions $B\parallel ab$ and $B\parallel c$. This is as expected for a free ion with $S=7/2$. However, for magnetic fields below 4 T, a slightly anisotropic behavior is observed. The measured magnetization in the $ab$-plane is larger than the magnetization along the $c$ axis, indicating a weak easy plane anisotropy. These observations are consistent with the previous studies~\cite{SG2019}. However, to explore the effect of the geometrical frustration, characterizations at lower temperatures are needed.

\begin{figure*}[ht!]
\includegraphics[width=6.8in]{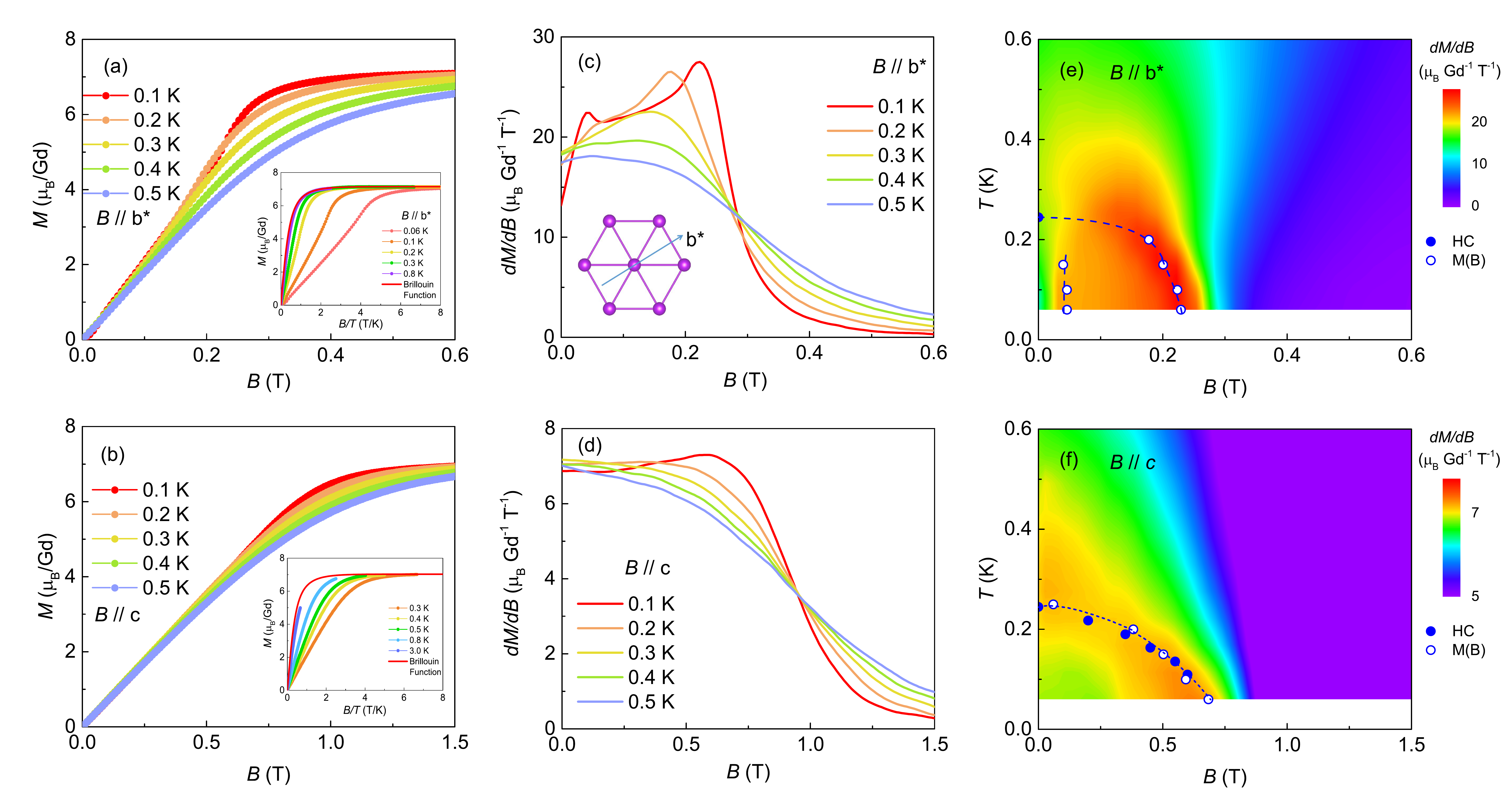}
    \caption{(a) and (b) The field dependent magnetization $M$ measured at different temperatures in $b^*$ and $c$ directions, respectively. Insets: Comparison of the the measured magnetization and the simulated Brillouin function of $S=7/2$.  (c) and (d) The field derivative magnetic susceptibility $dM/dB$ at different temperatures in $b^*$ and $c$ directions, respectively. (e) and (f) The field-temperature magnetic phase diagram overlaid on the contour plots of magnetic susceptibility $dM/dB$ with field along the $b^*$ and $c$ directions, respectively. }
\label{MH}
\end{figure*}

To further investigate the ground state, magnetization measurements down to very low temperatures were performed, using a home-built Hall sensor magnetometer integrated with the dilution refrigerator probe~\cite{MA1084,AC2004,AC2006}. For field in the $ab$-plane, the (110) direction (or $B\parallel b^*$) perpendicular to the hexagonal crystal edge was chosen. The isothermal magnetization $M(B)$ curves measured at different temperatures are presented in Fig.\ref{MH}(a) and (b), respectively. When the measurement temperatures are well above the ordering temperature $T_{\rm N}=0.24$ K, the field-dependent magnetization follows the Brillouin function (insets of Fig.\ref{MH}(a)and (b)). Upon lowering the temperature, these magnetization data start to deviate from the paramagnetic behavior. As evidenced by the field derivative magnetic susceptibility $dM/dB$, sharp peaks built up as the temperature is reduced below $T_{\rm N}$ (Fig.\ref{MH}(c) and (d)). Finally, two peak-like anomalies are observed at about 0.03 T and 0.24 T, at 60 mK for $B\parallel b^{*}$. The magnetic phase diagram for $B\parallel b^*$ is summarized in Fig.\ref{MH}(e). The intensity of the contour plot corresponds to the values of $dM/dB$. This phase diagram is similar to the observations in the Gd-based perovskites, such as GdAlO$_3$~\cite{KW1968} and GdScO$_3$~\cite{JM2020}. With only the nearest-neighbor interactions considered, the magnetic ground state of the easy plane triangular system was proposed to be a long-range ordered phase with spins oriented $120^{\rm o}$ to each other. For $B\parallel b^{*}$, the zero field $120^{\rm o}$ order is destroyed with increasing field, and spin flop-like transitions separate the magnetic phase diagram into two different regions. In the end, above the upper critical field at 0.24 T, all the Gd spins are fully polarized along the magnetic field.

On the contrary, for $B\parallel c$ only a single peak in $dM/dB$ is observed for a field around 0.7 T at 0.06 K. The magnetic phase diagram for the field along $c$ is presented in Fig.\ref{MH}(f)). In this case, the magnetic field is applied perpendicular to the Gd moments. The $120^{\rm o}$ ordered spins form a canted "V"-shaped phase, before fully polarizing along the $c$-axis. Since this is the hard direction, a larger saturation field is needed, compared to the saturation field 0.24 T for $B\parallel b^{*}$. It is worth to point out that isotropic behavior is expected if only the ground state $^8S_{7/2}$ with $L=0$ are considered. However, for Gd$^{3+}$ ions, it was found that the mixing of the higher-level states of $L\neq0$ cannot be neglected, especially for the physics at energy levels around $1\sim2$ K~\cite{Glazkov2005,Glazkov2006,Ali2011}. This could be the origin of the easy plane anisotropy observed in KBaGd(BO$_3$)$_2$. To clarify the single ion anisotropy, further studies such as electron spin resonance (ESR) measurements are still needed in the future.

\subsection{Specific Heat, Magnetic Entropy and Magnetocaloric Effect}
\begin{figure*}[ht!]
\includegraphics[width=6.8in]{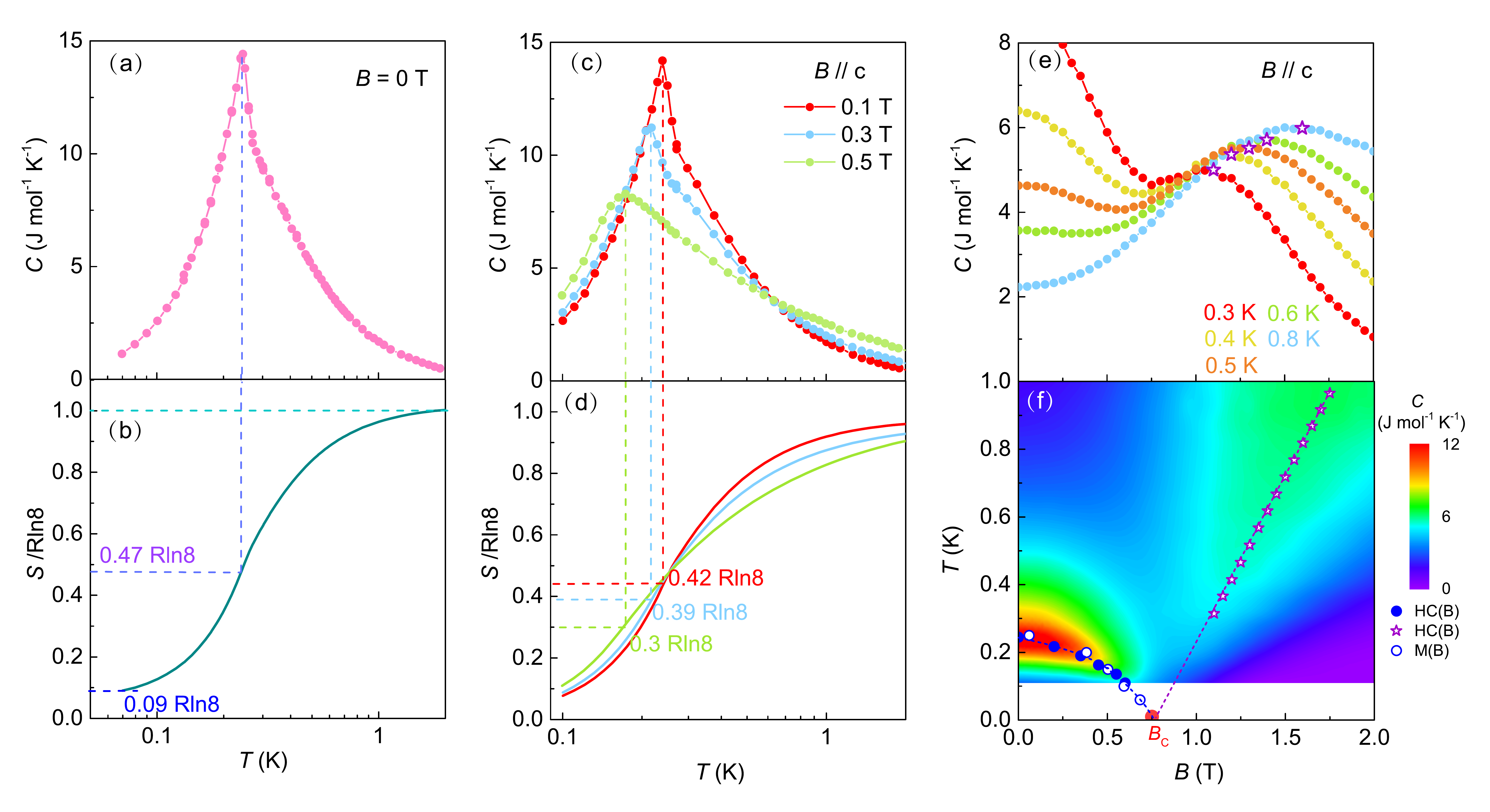}
    \caption{(a) The temperature dependence of the specific heat $C$ measured in zero fields. (b) The integrated magnetic entropy $S$ as the function of temperature. The vertical dashed line indicates the magnetic transition temperature $T_{\rm N}=0.24$ K. (c) The temperature dependence of the specific heat below the critical field along $c$ axis. (d) The temperature dependence of magnetic entropy below the critical field. (e) The field dependence of specific heat at different temperatures measured above the saturation fields along the $c$ axis. (f) The field-temperature magnetic phase diagram overlaid on the contour plots of specific heat in $c$ axis.}
\label{HC}
\end{figure*}

Specific heat measurements were performed as well. Shown in Fig.~\ref{HC}(a) is the temperature-dependent specific heat measured in zero field. A $\lambda$-shaped peak centered at $T_{\rm N}=0.24$ K is observed. This indicates the establishment of a long-range magnetic order. However, unlike other traditional magnets, a long tail extending to about 2 K is observed in the specific heat, which is about eight times higher than the transition temperature $T_{\rm N}=0.24$ K.  As also evidenced in the integrated magnetic entropy, only half of the full entropy ($0.47R\rm ln8$) is released at $T_{\rm N}=0.24$ K, and the full value of $R\rm ln8$ as expected for spins of $S$=7/2 is realized above 2 K. To account for the contribution below 70 mK, an estimated residual entropy with $S_0\simeq0.09R \rm ln8$ is added. These results are in contrast to the observations found in classical Gd-based magnets, such as GdAlO$_3$~\cite{Mahana2017}. In the case of GdAlO$_3$, almost the full entropy of $R\rm ln8$ was found at the transition temperature, and only very weak contribution was left above the transition temperature.  On the other hand, phenomena similar to what we find here, have been commonly observed in low dimensional quantum magnets~\cite{Wu2019A}, where large spin fluctuations are usually present.


Additional specific heat under field has been measured and shown in Fig.~\ref{HC}. Due to the thin flat plate-shaped crystal geometry, only data with field along the $c$-axis were collected. From these temperature dependent specific heat measured at different magnetic fields, it is noticed that the peak positions, together with the peak values in specific heat, are gradually suppressed with increasing fields(Fig.~\ref{HC}(c)). As evidenced in the integrated entropy (Fig.~\ref{HC}(d)), only about $0.3R\rm ln8$ is released at $T_{\rm N}$ at 0.5 T. All these phenomena indicate that a short range order persists in a wide temperature region above the transition temperature. This indicates that the presence of the spin frustrations in this two-dimensional triangular layer plays an important role.

\begin{figure*}[ht!]
\includegraphics[width=6.8in]{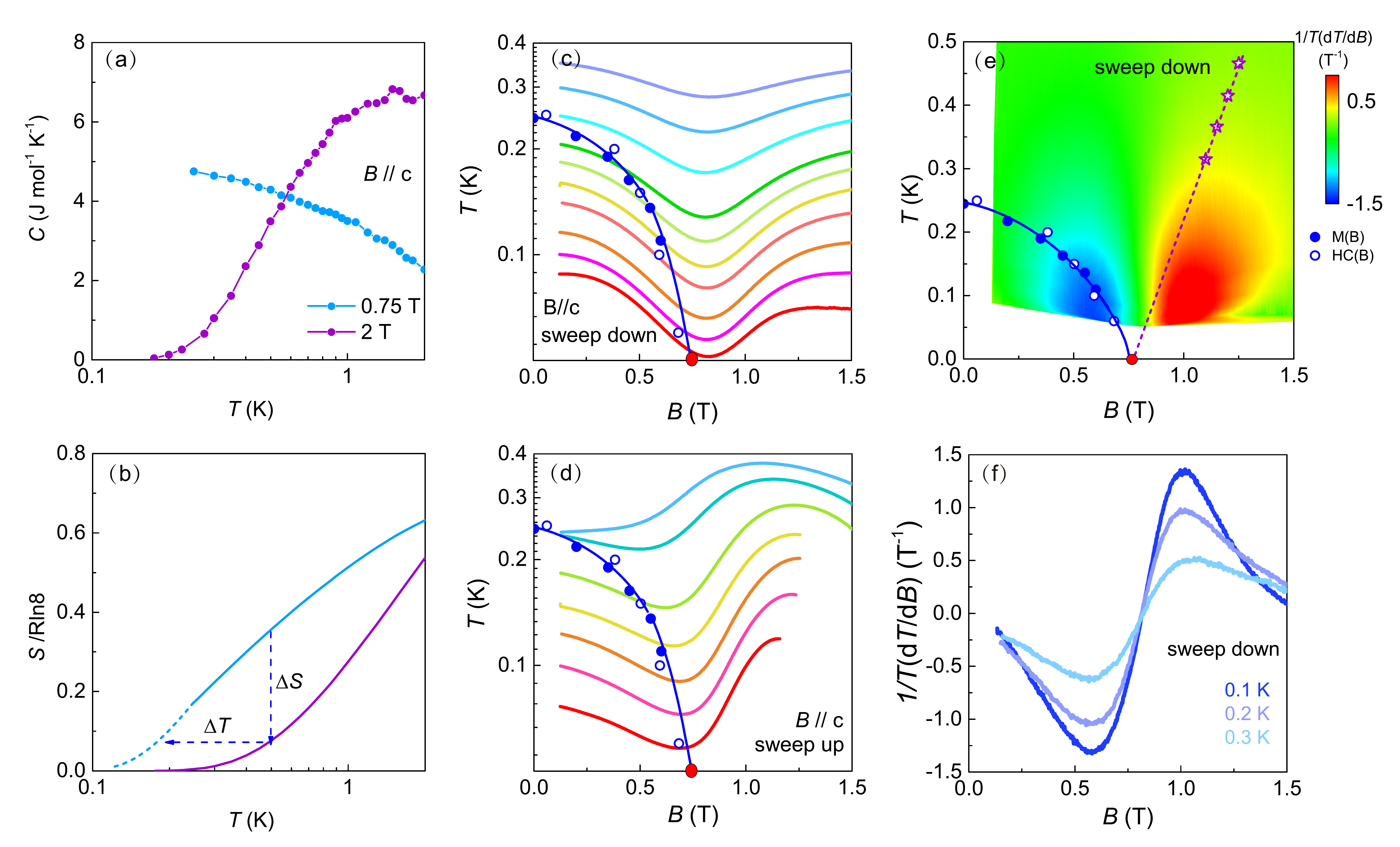}
    \caption{(a) The temperature dependence of the specific heat at and above the critical field along the $c$ axis. (b) The temperature dependence of the magnetic entropy at and above the critical field. Magnetocaloric effect of KBaGd(BO$_3$)$_2$ with the application of a magnetic field sweep down(c) and sweep up(d). (e) The contour plot of magnetic Gr\"{u}neisen parameter $\Gamma _{B}=1/T(dT/dB)$. (f) The magnetic Gr\"{u}neisen parameter $\Gamma _{B}$ as a function of magnetic field at several temperatures.}
\label{MCE}
\end{figure*}

Moreover, the antiferromagnetic transition is fully suppressed at the critical field at $B_{\rm c}\simeq0.75$ T. In the meantime, the measured specific heat monotonically increases with lower temperature (blue dots in Fig.~\ref{HC}(e)), suggesting that a significant residual entropy is pushed to zero temperature. As we continue to increase the field above $B_{\rm c}$, a broad peak-like anomaly is observed at higher temperatures (purple dots in Fig.~\ref{HC}(e)). In addition, these peak positions (indicated as stars in Fig.~\ref{HC}(e)) consistently move to higher temperatures with increasing fields. We have over-plot all these specific heat data together with the magnetic phase diagram (Fig.~\ref{HC}(f)). It is noticed that these peak positions extend linearly to the critical field $B_{c}$, indicating a gapless quantum critical point. These crossover-like behaviors above $B_{c}$ are usually related to a gap re-opening beyond the critical field. Interestingly, these crossover temperatures usually scale with the power law $T*\propto (B-B_{c})^{\nu z}$, and the linearity suggests a two-dimensional universality of $v z=1$, similar to the recent observations found in the triangular $S=1/2$ system $\rm Na_2BaCo(PO_4)_2$~\cite{Sheng2022}.

Presented in Fig.~\ref{MCE}(a) and (b) are the specific heat and the integrated entropy at 0.75 T and 2 T. The measured specific heat at 0.75 T keeps increasing as $T\rightarrow0$, indicating a large enhancement of the zero-temperature residual entropy at the field induced critical point. Due to the large relaxation time, the specific heat below 0.2 K at at 0.75 T was hard to measure. The residual entropy was estimated to be about $0.2R\rm ln8$ by linearly extracting the specific heat $C/T$ to zero as approaching 0 K. Meanwhile, the specific heat at low temperatures are greatly suppressed at 2 T, above the critical fields. This large entropy change near the critical region can be used for adiabatic demagnetization refrigeration~\cite{Wolf2011}. As illustrated in Fig.~\ref{MCE}(b)(blue dashed lines), a significant temperature drop $\Delta T$ can be expected by lowering the magnetic field from 2 T to 0.75 T due to the entropy change $\Delta S$. To test this scenario, magnecaloric effect (MCE) measurements were performed on a single crystal of the mass about 0.1 mg  with $B\parallel c$.  As the external fields were swept up and down, the sample temperatures were simultaneously recorded (Fig.~\ref{MCE}(c) and (d)). Instead of a sharp $V$-shaped anomaly, a rounded minimum like feature was observed around the phase boundary. This is due to a small heat leaksge from the environment, and the sample is not in an ideal adiabatic condition. To achieve temperatures below 50 mK, measurements with larger crystals and better designed MCE pucks are required in the future. However, even with the current setup using such a tiny crystal, we can still see a great magnetocaloric effect with a large magnetic Gr\"{u}neisen parameter $\Gamma _{B}=1/T(dT/dB)$ at the critical point (Fig.~\ref{MCE}(c)). The amplitude of the magnetic Gr\"{u}neisen parameter $\Gamma _{B}$ increases with decreasing temperature near the critical region (Fig.~\ref{MCE}(c)). This behavior is quite different from those systems where only weak fluctuations persist. In this case, the MCE curve becomes almost flat, and the amplitude of the magnetic Gr\"{u}neisen parameter drops to zero as the temperature was suppressed to zero~\cite{Wu2011}. The enhanced MCE effect in KBaGd(BO$_3$)$_2$ makes it a distinct promising candidate for adiabatic demagnetization cooling below 100 mK. In addition, due to the large magnetic entropy arising from the large spin value $S=7/2$, the demagnetization cooling of KBaGd(BO$_3$)$_2$ at the critical field will be more efficient comparared to other $S=1/2$ based materials.


\section{Conclusions}
To summarize, detailed investigations with low-temperature magnetization, and specific heat have been performed on KBaGd(BO$_3$)$_2$ single crystals. The magnetic Gd$^{3+}$ ions form into a two-dimensional triangular lattice with a large spin $S=7/2$. As the temperature is lowered, a long-range magnetic order is established below $T_{\rm N}=0.24$ K. To explore this ordered ground state, magnetic fields were applied in the triangular plane and along the $c$-axis. Although isotropic single ion properties are expected, the resulting magnetic phase diagrams are different from each other. The differences in the two phase diagrams arise from the weak easy plan anisotropy, and they are consistent with the $120^{\rm o}$ spin order in the plane. Thus, for field applied along the $b^{*}$ axis ($B\parallel(110)$), spin flop-like transitions with two successive critical fields are observed. While for field perpendicular to the magnetic moments with $B\parallel c$, the spins continuously rotated from the original (110) directions, until they are fully polarized along the $c$ axis at the saturation field. Although the large spin values of Gd$^{3+}$ ions do not make them as quantum as spin $S=1/2$, wide temperature regions of short range spin correlations are observed above the transition temperature, as evidenced by the specific heat and integrated magnetic entropy. This suggests that even for the spin magnitude as large as 7/2, the zero point fluctuations proportional to $1/S$ still play an important role as $T\rightarrow0$. On the other hand, these enhanced spin fluctuations and the large residual magnetic entropy hosted in KBaGd(BO$_3$)$_2$ can be utilized to realize efficient magnetic refrigeration in a moderate field region.

\begin{acknowledgments}
The research at SUSTech was supported by the National Key Research and Development Program of China (Grant No.~2021YFA1400400), the National Natural Science Foundation of China (Grants No.~12134020, No.~11974157, No.~12174175, and No.~12104255), the Guangdong Basic and Applied Basic Research Foundation (Grant No.~2021B1515120015), the Science, Technology and Innovation Commission of Shenzhen Municipality (No. ZDSYS20190902092905285). The Major Science and Technology Infrastructure Project of Material Genome Big-science Facilities Platform was supported by Municipal Development and Reform Commission of Shenzhen.
\end{acknowledgments}

\end{document}